\documentclass[aps,pra,twocolumn,superscriptaddress]{revtex4}
\usepackage{graphicx}
\usepackage{epsfig}
\usepackage{amsmath}
\usepackage{bm}
\usepackage{multirow}
\allowdisplaybreaks

\begin{document}
\title{Quantum phase transition of bosons in a shaken optical lattice}
\author{Jiao Miao}
\affiliation{Institute for Advanced Study, Tsinghua University, Beijing, 100084, China}
\author{Boyang Liu}
\affiliation{Institute for Advanced Study, Tsinghua University, Beijing, 100084, China}
\author{Wei Zheng}
\affiliation{Institute for Advanced Study, Tsinghua University, Beijing, 100084, China}

\begin{abstract}
Recently, the lattice shaking technique has been used to couple different Bloch bands resonantly. For the one-dimensional (1D) case, in which shaking is along only one direction, experimental observation of domain-wall formation has been explained by superfluid Ising transition. Inspired by these, we generalize to a 2D case in which shaking is along two orthogonal directions. Analogous to the 1D case, we find three different phases, the normal superfluid (NSF) phase, the $D_4$  symmetry-breaking superfluid ($D_4$SF) phase and the Mott insulator (MI) phase. Furthermore, we demonstrate that the interaction effect induced by inhomogeneous band mixing can modify the critical shaking amplitude. Unlike in the 1D case, shaking types also can modify the critical shaking amplitude.  Unlike in the 1D case, shaking types also can modify the critical shaking amplitude. We also construct a low-energy effective field theory to study the quantum criticality of bosons near the tricritical point of NSF,  $D_4$SF and MI phases.  Moreover, we find a Bose liquid with anisotropically algebraic order and propose to change the Bose-Einstein condensation (BEC) into a non-condensed Bose liquid by tuning the shaking amplitude approaching the critical value.
\end{abstract}

\maketitle

\section{introduction} \label{intro}

More and more interest has been attracted to ultracold atoms trapped in a time-periodically driven optical lattice. There are two cases, off-resonance and resonance. For the off-resonance case, it was stated that shaken lattice system can be described by an effective time-independent Hamiltonian with renormalized hopping amplitudes for a large shaking frequency  \cite{Holthaus}. It was also experimentally demonstrated that hopping amplitude can be changed dynamically with maintained phase coherence of condensation by shaking the lattices \cite{Lignier}. The lattice shaking technique can be used to tune hopping parameters and even invert the signs in a coherent way, which opens a new direction to simulate quantum phase transitions in ultracold atom systems.  Coherent control of the superfluid-Mott-insulator (-MI) phase transition has been realized in a shaken three-dimensional optical lattice  \cite{Zenesini}. A synthetic gauge field can be realized in a shaken optical lattice \cite{Hemmerich,Sengstock1,Sengstock2,Sengstock3}, and this is equivalent to insetting a $\pi$ flux in each plaquette in a shaken square lattice, which generates a staggered-vortex superluid state \cite{Hemmerich}, or in each triangle in a shaken triangular lattice, which generates  various types of frustrated states \cite{Sengstock1,Sengstock3}. In addition, interparticle interaction can be tuned from repulsive to attractive in fermionic lattice systems by ac forcing, which allows one to simulate an attractive Hubbard model effectively with temperatures below the superconducting transition temperature \cite{Tsuji}.

 The resonance case starts from experimental observation of domain-wall formation for bosons condensed in a shaken one-dimensional (1D) optical lattice \cite{Chin}, in which the lattice shaking technique hybrids different Bloch bands. The effective Hamiltonian cannot be described by renormalized hopping amplitudes or interactions as used in off-resonant cases, which may lead to novel phases. The finite-momentum superfluid phase with spontaneously broken $Z_2$ symmetry called the $Z_2$SF phase has been observed \cite{Chin} and the corresponding normal superfluid- (NSF-)$Z_2$SF-MI phase transition has been described by a low-energy effective field theory in the 1D case \cite{ising}.

The finite-momentum condensate has been realized by spin-orbit (SO) coupling generated by Raman transitions \cite{soc1,soc2}, or in a staggered magnetic field \cite{stagger}, or in a shaken optical lattice \cite{Chin,Hemmerich,Sengstock2}. The condensate with finite momentum has spatially inhomogeneous order parameter, which is a bosonic analogy to the Fulde-Ferrell-Larkin-Ovchinnikov phase in superconductors \cite{FFLO}. Inspired by the discovery of a finite-momentum condensate by resonantly shaking a lattice along one direction \cite{Chin,ising}, in this paper, we generalize to a two-dimensional (2D) case. Using Floquet theory, we demonstrate formally and numerically that lattice shaking leads to a phase transition from the NSF phase to the $D_4$SF phase as the shaking amplitude increases. We further show inhomogeneous band-mixing-induced interaction effect modifies the critical shaking amplitude, which is analogous with the 1D case \cite{ising}. There is a notable difference between our model in the 2D case with the model in the 1D case \cite{ising}. There are various shaking types. For example, a lattice can be shaken along one diagonal of the lattice (linear shaking), or elliptically (elliptical shaking), or circularly (circular shaking). Since separability of the system exists along two primitive vectors, quasienergy dispersion is independent of shaking types. However, linear shaking preserves time-reversal (TR) symmetry, while elliptical or circular shaking breaks TR symmetry. Analogous to orbital Hund's rule \cite{hund}, there is the largest interaction energy at fixed momentum for repulsive bosons with linear shaking than for other shaking types. Together with inhomogeneous band mixing, we predict the smallest critical shaking amplitude for linear shaking. Then we construct a low-energy effective theory to describe phase transitions. A critical correlation length exponent is calculated by the momentum shell renormalization-group (RG) method. In the end, we study the existence of a Bose-Einstein condensate (BEC) in a general shaken lattice system. A lot of effort has been devoted to realizing quantum states which are not Bose condensed \cite{bose liquid1,bose liquid2,bose liquid3,bose liquid4,bose liquid5,bose liquid6,bose liquid7,bose liquid8,bose liquid9,bose liquid10,bose liquid11,bose liquid12}. We find a Bose liquid with an anisotropically algebraic order in a three-dimensional lattice with two directions shaken, and we propose to change the BEC into a noncondensed Bose liquid via tuning shaking amplitude approaching the critical value. 

The paper is organized as follows. In Sec. \ref{model}, we introduce the model for bosons in a shaken optical lattice. In Sec. \ref{finite momentum SF}, we calculate the quasienergy spectrum and obtain a finite-momentum superfluid phase.  Next, we study the interaction effect on this phase in Sec. \ref{D4SF} and \ref{fc}. In Sec. \ref{field theory}, we construct a low-energy effective field theory to study the quantum criticality of the phase transition. The existence of a BEC in a general shaken lattice system is discussed in Sec. \ref{non condensate}. Finally, conclusions are presented in Sec. \ref{conclusion}.

\section{MODEL} \label{model}
The system we consider is two counter-propagating laser beams along the $x$ direction and two along the $y$ direction, which forms a square lattice. The lattice is shaken by time-periodically modulating relative phase $\varphi_x(t)$ between laser beams along the $x$ direction and $\varphi_y(t)$ between that along the $y$ direction via acousto-optic modulators. The Hamiltonian reads  
\begin{equation}
\hat{H}(t)=\frac{\hat{\bm{p}}^2}{2m}+V \cos^2 (k_r x+\frac{\varphi_x(t)}{2} )+ V \cos^2 (k_r y+ \frac{ \varphi_y(t)}{2} ),  \label{H}
\end{equation}
where $k_r$ is photon momentum, $\varphi_x(t)=f \cos \omega t, \varphi_y(t)=f \cos (\omega t+\varphi)$, $f$ is the shaking amplitude, and $\varphi$ is the relative phase between $\varphi_{x}$ and $\varphi_y$. $\varphi=0$ or $\pi$ means  preserving TR symmetry, while $\varphi \neq 0$ and $\pi$ means breaking TR symmetry. $\Delta \equiv f/ (2 k_r )$ is the maximum lattice displacement along the $x$ or $y$ direction.  This model is separable along the $x$ and $y$ direction. 

Taking a transformation $x \rightarrow x - \Delta  \cos \omega t$, $y \rightarrow y - \Delta  \cos (\omega t+\varphi)$, the Hamiltonian in the comoving frame reads 
\begin{align}
\hat{H}(t,\varphi)&=\frac{1}{2m} \hat{\bm{p}}^2+V \cos^2 (k_r x) + V \cos^2 (k_r y) - \frac{ \bm{A}(t) \cdot \hat{ \bm{p} } }{ m } ,  \label{Ht}
\end{align}
where the effective vector potential  is $\bm{A}(t)=m \omega \Delta  (\sin \omega t, \sin (\omega t+\varphi) )$. Neutral particles will act as charged particles in a static square lattice and an ac electric field $\bm{E}= - m \omega^2 \Delta  (\cos \omega t, \cos (\omega t+\varphi))$. The effective charge is set to be unity. 

The first three static terms in Eq.(\ref{Ht}) give a static band structure $\epsilon_{\lambda}(\bm{k})$ and corresponding Bloch wave function $\phi_{\lambda, \bm{k}}(\bm{r})$, which will serve as basis in the following analysis. In this paper, we consider shaking frequency $\omega$ is a little blue-detuned from $p_x$ and $p_y$ bands. Moreover, we notice higher bands couples with $s$-band via higher-order processes due to symmetry and hence only keep $s$, $p_x$ and $p_y$ bands. And we numerically verify that our following qualitative results do not change when counting higher bands.  In these bases, the tight-binding form of the Hamiltonian in the comoving frame is given by
\begin{align}
\hat{H}(t,\varphi)=\sum_{\bm{k}} 
\left( 
\hat{\Psi}^{\dagger}_{p_x, \bm{k}}, \hat{\Psi}^{\dagger}_{p_y,\bm{k}}, \hat{\Psi}^{\dagger}_{s, \bm{k}}
\right) 
H_{\bm{k}}(t,\varphi)
\left(\begin{array}{c}
\hat{\Psi}_{p_x, \bm{k}} \\
\hat{\Psi}_{p_y, \bm{k}} \\
\hat{\Psi}_{s, \bm{k}}
\end{array} \right),
\end{align}
where $\hat{\Psi}^{\dagger}_{\lambda, \bm{k}}$ and $\hat{\Psi}_{\lambda, \bm{k}}$ are creation and annihilation operators of a particle with quasimomentum $\bm{k}$ in $\lambda$ band, respectively, and $\lambda$ is $p_x,p_y$ or $s$. 

\begin{figure}[tbp]
\includegraphics[height=2.8in, width=2.8in]
{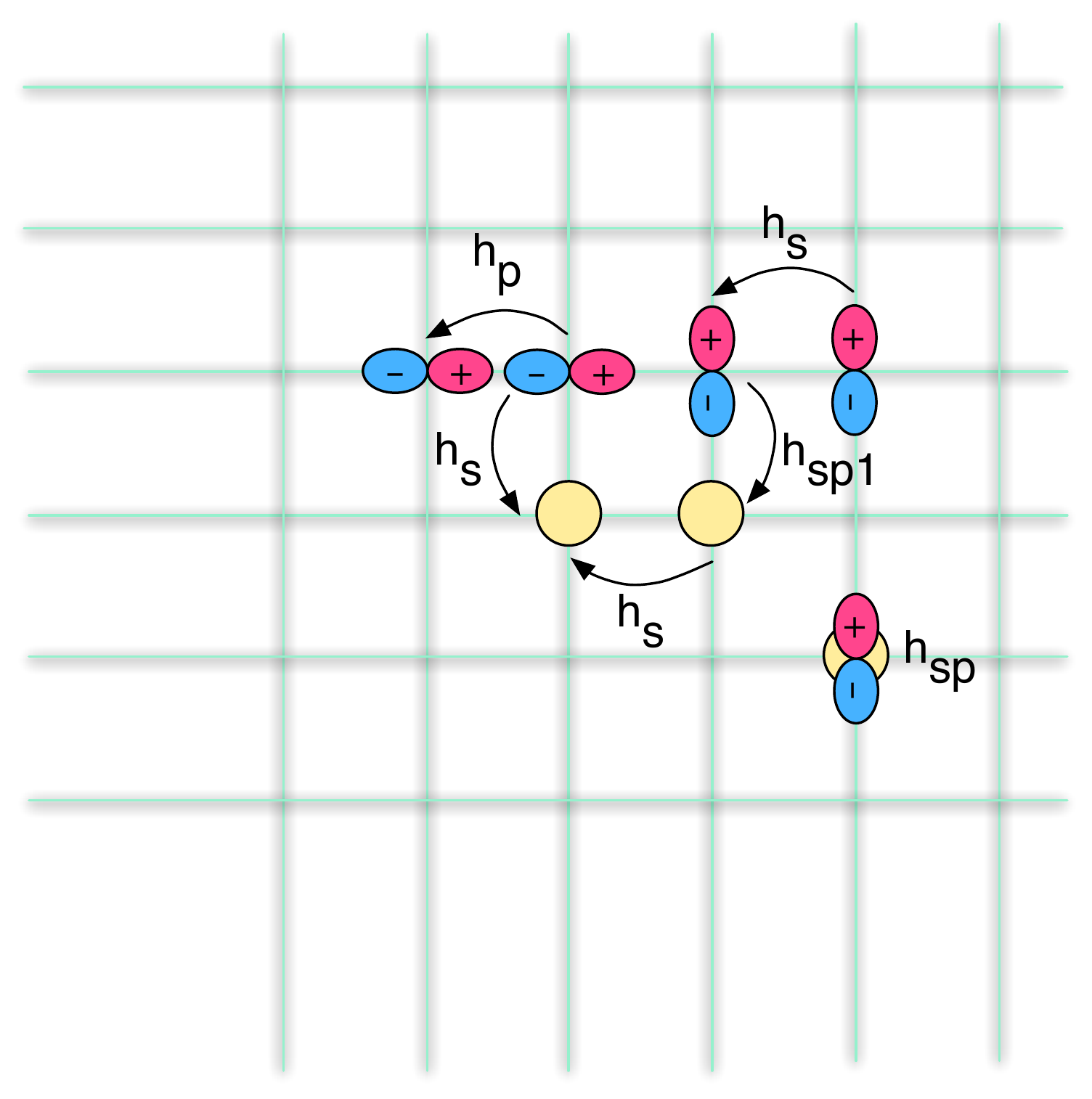}
\caption{Shaking induced couplings in real space. The circles (yellow) denote $s$-orbital, the colored shapes laid along the horizontal direction and that laid along the vertical direction denote the $p_x$ and $p_y$ orbitals, respectively, and $+$ and $-$ signs denote the orbital phase.}
\label{coupling}
\end{figure}

The Hamiltonian in momentum space is given by 
\begin{align}
 H_{\bm{k}}(t,\varphi) 
&=\left(\begin{array}{ccc}
\epsilon_{p_x}(\bm{k})   &   0   &  0 \\
0 & \epsilon_{p_y}(\bm{k}) & 0 \\
0 & 0& \epsilon_s (\bm{k})
\end{array}\right)  - \frac{ A_x(t) }{ m } \times 
 \nonumber 
 \\ 
& \quad 
\left(\begin{array}{ccc}
2 h_p  \sin k_x   &   0   & - i \Omega(k_x) \\
0 & 2 h_s  \sin k_x & 0 \\
i \Omega(k_x) & 0 & 2 h_s  \sin k_x
\end{array}\right)   - \frac{ A_y(t) }{ m } \times
 \nonumber 
 \\
&  \quad  
\left(\begin{array}{ccc}
2 h_s  \sin k_y   &   0   &  0 \\
0 & 2 h_p  \sin k_y &  -i \Omega(k_y) \\
0 &  i \Omega(k_y) & 2 h_s  \sin k_y
\end{array}\right) ,     \label{H_tbm}
\end{align}
where
\begin{align}
\Omega(k_{x,y})&=h_{sp}+2 h_{sp1} \cos k_{x,y} , 
\\
h_s&=  \langle w_{s,p_y}(\bm{r})| i \hat{p}_x |w_{s,p_y}(\bm{r}-\bm{e}_x) \rangle  
 \\
h_p&=   \langle w_{p_x}(\bm{r})| i \hat{p}_x |w_{p_x}(\bm{r}-\bm{e}_x)\rangle, 
\\
h_{sp}&=  \langle w_{p_x}(\bm{r})| i \hat{p}_x|w_{s}(\bm{r})\rangle,  
\\
h_{sp1}&=  \langle w_{p_x}(\bm{r})| i \hat{p}_x |w_{s}(\bm{r} - \bm{e}_x)\rangle ,
\end{align}
where $A_{\gamma}(t)$ is the $\gamma$ component of the effective vector potential $\bm{A}(t)$, $\bm{e}_{\gamma}$ is the primitive vector along the $\gamma$ direction, $\gamma$ is $x$ or $y$, $w_{\lambda}(\bm{r})$ is the Wannier function of the $\lambda$ band, $\lambda$ is $p_x$, $p_y$ or $s$, and $\langle \cdots | \cdots | \cdots \rangle$ denotes integral in the coordinate space $\int d \bm{r} \cdots $. Real coupling amplitudes $h_s, h_p, h_{sp1}, h_{sp}$ denote shaking induced nearest-neighbor hopping between $s$ bands,  between $p$ bands, between $s$ and $p$ bands, and onsite coupling between $s$ and $p$ bands, respectively, as shown in Fig. \ref{coupling}. 
The first matrix in Eq.(\ref{H_tbm}) represents a static band structure. And the last two matrices represent shaking-induced coupling among three bands. It is essential that lattice shaking induces hopping between $s$ and $p$ bands, which is symmetry forbidden in the absence of shaking. Here shaking plays the role of external field breaking inversion symmetry, which is similar to mixing the $p_z$ band with $p_{x,y}$ band by an external electric field in the orbital Rashba effect \cite{Han}.

\begin{figure}[tbp]
\includegraphics[height=2in, width=3.5in]
{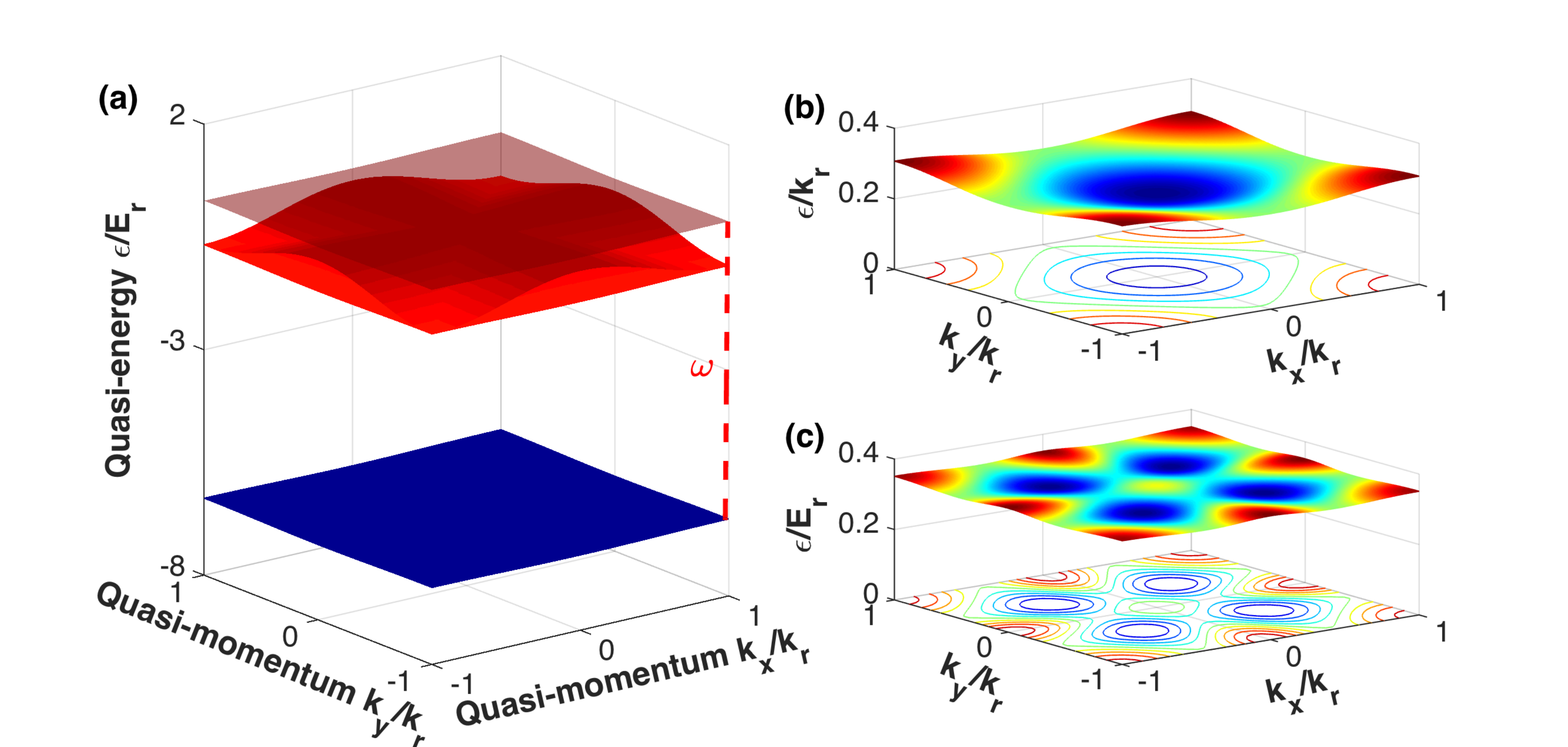}
\caption{ Quasienergy spectrum with $V=13E_r, \hbar \omega=6.6E_r, \varphi=0$, where $ E_r = \hbar^2 k_r^2 / ( 2 m ) $ is the lattice recoil energy. (a) Quasienergy spectrum before shaking. The half transparent surface denotes the dressed $s$ band with energy lifted by $ \hbar \omega $. (b,c) Quasienergy dispersion of the uppermost band for (b) $f=0.02$ and (c) $f=0.08$. }
\label{non-int}
\end{figure}

\section{FINITE-MOMENTUM PHASE}  \label{finite momentum SF}

By diagonalizing the Floquet operator, i.e., time-revolution operator in a time period $T_0=2 \pi/\omega$,
\begin{equation}
 \hat{U}(T_0,\varphi)=\hat{\mathcal{T}} e^{-i \int_{0}^{T_0} \mathrm{d} t \hat{H}(t,\varphi)} ,
 \end{equation}
 one can obtain a quasienergy spectrum as shown in Fig. \ref{non-int}. There is a certain critical shaking amplitude $f_c^0$ such that for $f<f_c^0$, the uppermost band exhibits a single minimum at zero momentum, and for $f>f_c^0$, the uppermost band exhibits four minima at finite momenta. 
  
Symmetry of a periodically driven system must be considered at the Floquet operator $ \hat{U}(T_0,\varphi) $ level \cite{topo classification}. Since $\hat{U}(T_0,\varphi)$ is separable along $x$ and $y$ direction, quasienergies do not depend on the relative phase $\varphi$. For $\varphi=0$, the original Hamiltonian $\hat{H}(t,\varphi=0)$ in Eq. (\ref{Ht}) has $D_4$ symmetry, so does the quasienergy spectrum. So the uppermost band dispersion has $D_4$ symmetry for  any $\varphi$. 

To describe the system, an effectively static Hamiltonian $\hat{H}_{eff}$ is defined as
\begin{equation}
\hat{U}(T_0,\varphi) \equiv e^{-\frac{i}{\hbar} \hat{H}_{eff}(\varphi) T_0}.    \label{U}
\end{equation}
We will analyze a rotating-wave-approximation (RWA) Hamiltonian, i.e., the zero order $(1/ \omega)^0$ term of the effective Hamiltonian \cite{floquet topo}, which is given by
\begin{align}
H_{\bm{k}}^{RWA}(\varphi)&=\left( O^{\dagger}(t) (H_{\bm{k}}(t)-i \partial_t) O(t) \right)^{(0)} \nonumber 
\\
&=\left(\begin{array}{ccc} 
\epsilon_{p_x}(\bm{k})  &  0 & \tilde{\Omega} (k_x)   \\
0  &   \epsilon_{p_y}(\bm{k})   & e^{- i \varphi} \tilde{\Omega} (k_y)     \\
\tilde{\Omega}(k_x)    & e^{i \varphi} \tilde{\Omega}(k_y)   &  \epsilon_s(\bm{k})+\omega 
\end{array} \right),  \label{Hrwa}
\end{align}  
 where
\begin{align}
\tilde{\Omega}(k_{x,y}) & = - \frac{\omega \Delta}{2} \Omega (k_{x,y}),
\\
O(t) &= \left(\begin{array}{ccc}
1 & 0 & 0 \\
0 & 1 & 0 \\
0 & 0 & e^{i \omega t}  
\end{array}\right)   \label{rwa}
\end{align}
and superscript $(0)$ denotes the static part. Coupling strength $\tilde{\Omega}(k_{x,y})$ is proportional to shaking frequency and amplitude. The RWA Hamiltonian in Eq.(\ref{Hrwa}) indicates that lattice-shaking-induced couplings result in level repulsion effect, which is the strongest along $k_x=0$ and $k_y=0$ directions. The level repulsion effect combined with $D_4$ symmetry will give rise to four global minima at finite momenta $(\pm k_c, \pm k_c)$ instead of one at zero momentum in the uppermost band as shaking amplitude $f$ increases. 

The finite-momentum BEC has been proposed in a non-separable square lattice subjected to the off-resonant shaking \cite{finite momentum}. Our model has more orbital physics, which will be shown in Sec.\ref{fc}.

\section{SPONTANEOUS SYMMETRY BREAKING}  \label{D4SF}
Let us consider interacting bosons condensing at the finite-momentum state with minimal kinetic energy in the uppermost band. The interaction reads
\begin{equation}
H_{int}(t)= g \int \mathrm{d} \bm{r} \hat{\Psi}^{\dagger}(\bm{r},t)\hat{\Psi}^{\dagger}(\bm{r},t)\hat{\Psi}(\bm{r},t)\hat{\Psi}(\bm{r},t),
\end{equation}
where $\Psi^{\dagger}(\bm{r},t)$ and $ \Psi(\bm{r},t)$ are creation and annihilation operators of the condensate state, respectively, and positive $g$ is the repulsive interaction strength.  

Bosons can either condense at one of the four degenerate finite-momentum states or at the superposition state. Assume the single-particle ground state is a superposition state
\begin{equation}
\psi_{k_c}(\bm{r},t)=\sum_{i=1}^4 a_i \psi_{\bm{k}_i}(\bm{r},t),
\end{equation}
where $\psi_{\bm{k}_i}(\bm{r},t) $ is one of the four degenerate states, $\bm{k}_i=(\pm k_c, \pm k_c)$ is the condensate momentum, and constant $a_i$ satisfies $\sum_{i=1}^{4} |a_i|^2=1$. One can write $ \psi_{\bm{k}_i}(\bm{r},t) $ in the comoving frame as
\begin{align}
\psi_{\bm{k}_i}(\bm{r},t)&=b_{p_x,k_c} \phi_{p_x,\bm{k}_i}(\bm{r})+b_{p_y,k_c} \phi_{p_y,\bm{k}_i}(\bm{r}) \nonumber \\
& \quad + e^{i \omega t} b_{s,k_c} \phi_{s,\bm{k}_i}(\bm{r}),
\end{align}
where $b_{\lambda,k_c}$ is a combination coefficient and dependent on $\lambda$ and $k_c$, and $\lambda$ is $p_x, p_y$, or $s$.

The time-average mean-filed interaction energy per particle condensing at the superposition state $\psi_{k_c} (\bm{r},t)$ in the laboratory frame is given by
\begin{equation}
\epsilon_{int}(k_c)=\frac{1}{T_0} \int_0^{T_0} \mathrm{d} t g \int \mathrm{d} \bm{r} |\psi_{k_c} (\bm{r},t)|^4.    \label{int}
\end{equation}
 By minimizing the interaction energy $\epsilon_{int}(k_c)$ with respect to $\{a_i \}_{i=1}^4$, one obtains $(a_1, a_2, a_3, a_4)=(\pm 1, 0, 0, 0)$ or $(0, \pm 1, 0, 0)$ or $(0, 0, \pm 1, 0)$ or $(0, 0, 0, \pm 1)$. Since there is only one nonzero $a_i$, we neglect the phase of $ a_i $. So bosons only condense at one of the four finite-momentum states, which breaks $D_4$ symmetry spontaneously. 

In the process of turning on shaking adiabatically, bosons will remain in the uppermost band. When shaking amplitude across a critical value, phase transition from the NSF phase to the $D_4$SF phase happens.

\section{critical shaking amplitude}    \label{fc}
The interaction effect also modifies the critical shaking amplitude. By minimizing the total energy that consists of kinetic and interaction energies of bosons in the uppermost band with respect to quasimomentum, one can obtain the condensate momentum $( \pm k_c, \pm k_c)$. Here the methods we use to calculate kinetic and interaction energies are the same as the methods used in Sec. \ref{finite momentum SF} and \ref{D4SF}, respectively. When $k_c$ turns out to be nonvanishing with the increasing shaking amplitude, critical shaking amplitude $f_c$ is obtained. 

\begin{figure}[tbp]
\includegraphics[height=2in, width=3.5in]
{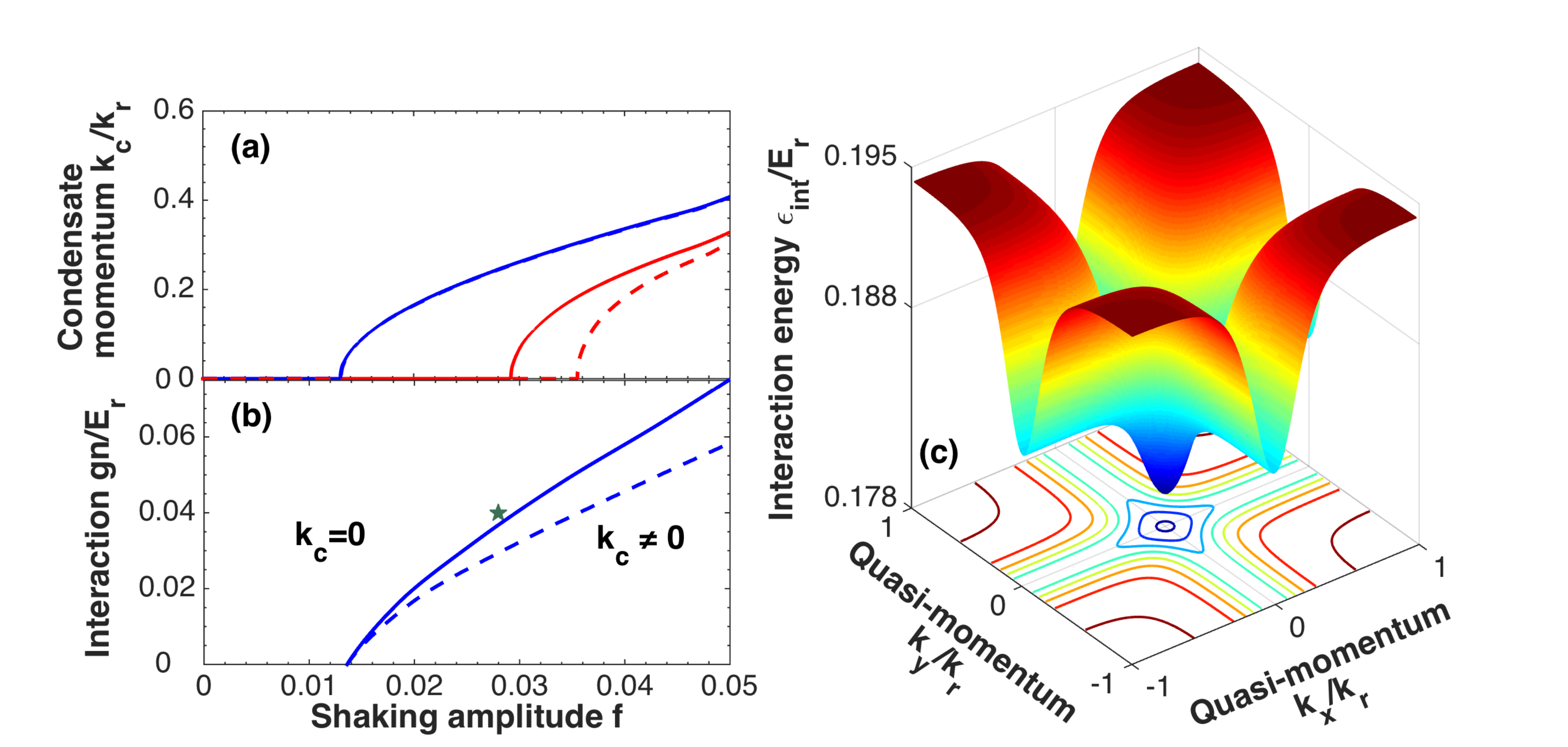}
\caption{(a,b) Interaction shifts of critical shaking amplitude with $V=13E_r, \omega=6.4E_r$. The solid line denotes $\varphi=0$ or $\pi$, and dashed line denotes $\varphi=\pm \pi/2$.  (a) Condensate momentum component $ k_c $ as a function of shaking amplitude $f$ for  non-interacting (blue), and interacting (red) cases with $ gn=0.04E_r$. (b) Phase diagram for a given shaking frequency $\omega$. Left region: NSF phase. Right region: $D_4$SF phase. (c) Interaction energy ($\epsilon_{int}$) with $\varphi=0$ and $(f, gn)$ marked as a star in (b).}
\label{k-f}
\end{figure}

Fig. \ref{k-f}(a,b) show repulsive interaction effect enlarges the critical shaking amplitude $f_c$ in the deep lattice limit. This is because inhomogeneous band mixing in momentum space causes a global minimum of interaction energy at zero momentum in the deep lattice as shown in Fig. \ref{k-f}(c), which has been illustrated in the 1D case \cite{ising}. Instead, a local maximum has also been predicted at zero momentum in the shallow lattice limit, which leads to a smaller critical shaking amplitude $f_c < f_c^0$ \cite{ising}. We will focus on the deep lattice case in this paper.

 Besides, Fig. \ref{k-f}(a,b) also show shaking types can modify critical shaking amplitude $f_c$. The $k_c-f$ curve for the interaction case in Fig. \ref{k-f}(a) and the $gn-f_c$ curve in Fig. \ref{k-f}(b) change with relative phase $\varphi$ and are bounded by corresponding curves with $\varphi=0$ or $\pi$ and that with $\varphi=\pm \pi/2$. 

 One can write the eigenstate of the uppermost band in the comoving frame as
\begin{align}
\psi_{\bm{k}}(\bm{r},t,\varphi)&=c_{p_x,\bm{k}} \phi_{p_x,\bm{k}}(\bm{r})+e^{-i \varphi}c_{p_y,\bm{k}} \phi_{p_y,\bm{k}}(\bm{r})  \nonumber 
\\
& \quad +e^{i \omega t} c_{s,\bm{k}} \phi_{s,\bm{k}}(\bm{r}),  \label{psi}
\end{align}
where $c_{\lambda,\bm{k}}$ is the  combination coefficient of the eigenstate of $H_{\bm{k}}^{RWA}(\varphi=0)$ in the uppermost band and $\lambda$ is $ p_{x}, p_y$ or $s$. Since $H_{\bm{k}}^{RWA}(\varphi=0)$ is real and symmetric, $c_{\lambda,\bm{k}}$ can be set to be real. Here we use the RWA Hamiltonian for simplicity of analysis.

Using  Eq. (\ref{int}), the difference between time-average interaction energy per particle in a system with broken TR symmetry and that in a system with TR symmetry is given by
\begin{align}
\Delta \epsilon_{int}(\bm{k},\varphi) & \equiv \epsilon_{int}(\bm{k},\varphi)-\epsilon_{int}(\bm{k},\varphi=0)
\nonumber \\
&= -4 \nu \sin^2 \varphi \, c_{p_x,\bm{k}}^2 c_{p_y,\bm{k}}^{ 2} U^{p_x p_y}_{\bm{k}}, 
\label{delta int}
\end{align}
where $\nu$ is the site occupation number and $U_{\bm{k}}^{p_x p_y} = g \int \text{d} \bm{r} |\phi_{\bm{p_x,k}}(\bm{r}) \phi_{\bm{p_y,k}}(\bm{r})|^2$.

Eq.(\ref{delta int}) shows two important ingredients. One is $-\sin^2 \varphi$  signifying level of TR symmetry breaking.

Generally, when one Hamiltonian with TR symmetry is unitarily transformed to another Hamiltonian, the spectrum rather than TR symmetry is always invariant. Assuming $\hat{\tilde{H}}=\hat{Q}^{\dagger} \hat{H} \hat{Q}$ with TR symmetric Hamiltonian $\hat{H}$ and unitary operator $\hat{Q}$, one obtains
\begin{equation}
[\hat{T}, \hat{\tilde{H}}]=( \hat{\tilde{Q}}^{ \dagger} \hat{H} \hat{\tilde{Q}}-\hat{Q}^{\dagger} \hat{H} \hat{Q} ) \hat{T},     \label{TR}
\end{equation} 
where $\hat{T}$ is the TR operator and $\hat{\tilde{Q}}=\hat{T}\hat{Q} \hat{T}^{-1}$. If $\hat{Q}$ commutes with $\hat{T}$, then $\hat{\tilde{Q}}=\hat{Q}$ and $\hat{\tilde{H}}$ has TR symmetry. Otherwise TR symmetry is broken generally. In our case, $\hat{H} \simeq \hat{H}_{\bm{k}}^{RWA}(0)$ has TR symmetry, $ \hat{Q} $ has the matrix form
\begin{align}
 \left(
\begin{array}{ccc}
1 & 0 & 0 \\      
0 & e^{ i \varphi} & 0 \\
0 & 0 & 1
\end{array}
\right),
\end{align}
and commutation relation $[\hat{T}, \hat{H}_{\bm{k}}^{RWA}(\varphi)]$ is proportional to $\sin \varphi$. So TR symmetry is conserved for $\varphi=0$ or $\pi$ and broken maximumly for $\varphi=\pm  \pi/2$. $\Delta \epsilon_{int} ( \bm{k}, \varphi ) $ in Eq.(\ref{delta int}) at the fixed momentum decrease as the level of TR symmetry breaking increases, which is similar to orbital Hund's rule \cite{hund}. 

\begin{figure}[tbp]
\centering
\includegraphics[height=2.5in,width=2.5in]
{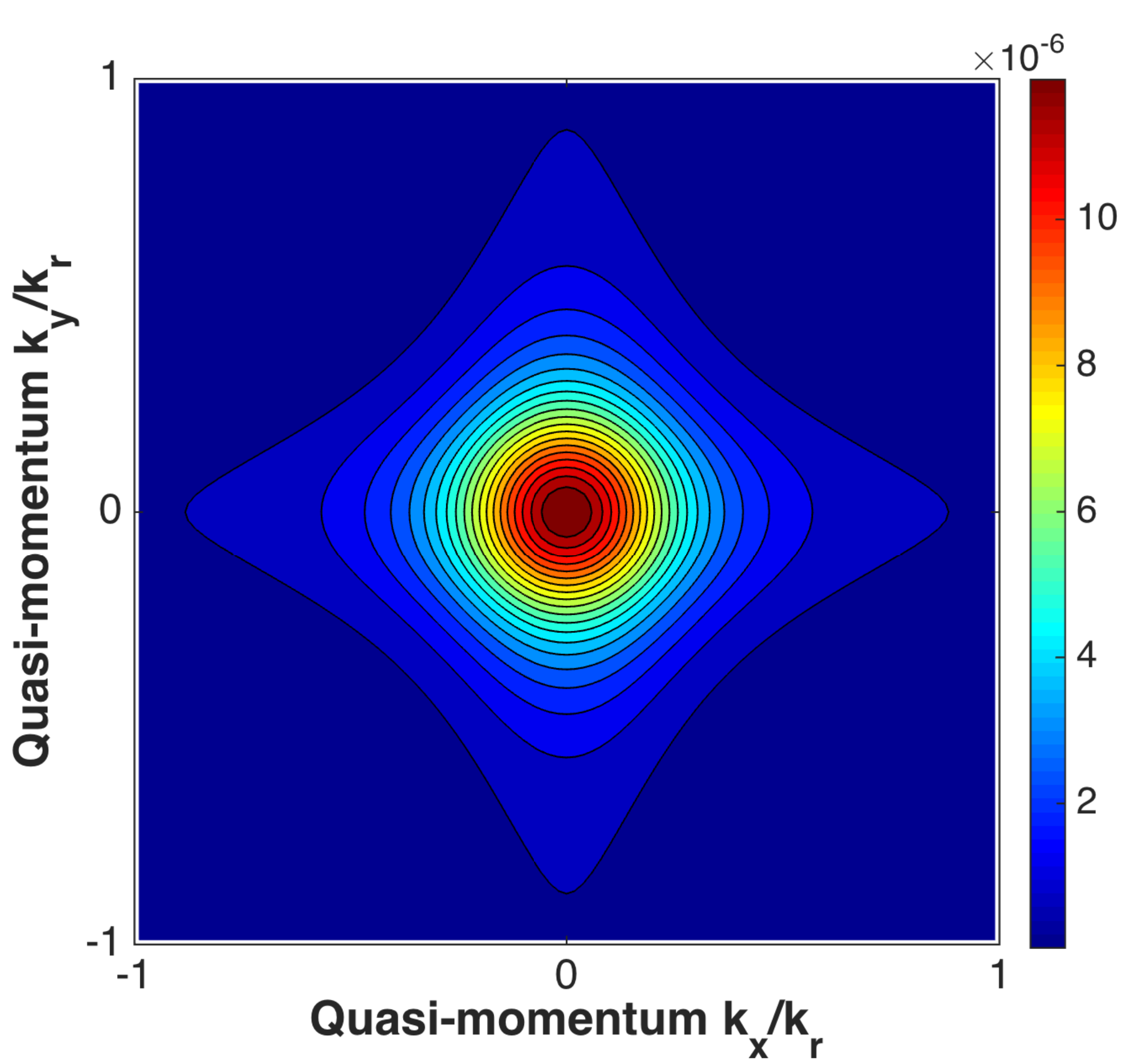}
\caption{Contour plot of $c_{p_x,\bm{k}}^2 c_{p_y,\bm{k}}^2 U_{\bm{k}}^{p_x,p_y} $, the momentum-dependent part of the interaction energy difference $ \Delta \epsilon_{int}(\bm{k},\varphi) $, with parameters same as used in Fig. \ref{k-f}(c). }
\label{inhomogeneous hybridization}
\end{figure}

The other key ingredient of interaction energy difference in Eq. (\ref{delta int}) is momentum dependence. $ U_{\bm{k}}^{p_x,p_y} $ is a positive constant in the deep lattice limit. Inhomogeneous band mixing in momentum space causes the fact that the momentum-dependent part $c_{p_x,\bm{k}}^2 c_{p_y,\bm{k}}^2 U_{\bm{k}}^{p_x,p_y} $ has a global maximum at zero momentum, as shown in Fig.(\ref{inhomogeneous hybridization}). Here we can see $\varphi$ appears only in interaction components involving $p_x$ and $p_y$ orbitals. The reason is that other terms involving $\varphi$, such as $c_{s,\bm{k}}^{ 2} c_{p_y,\bm{k}}^2 e^{\pm i 2 \varphi}$, contain factor $e^{\pm i 2 \omega t}$ because of the energy difference between $s$ and $p$ bands and can be neglected in the sense of time average. 

The two ingredients determine curvature of interaction energy at zero momentum is minimum when $\varphi=0$ or $\pi$ and maximum when $\varphi= \pm \pi/2$. So there is the smallest critical shaking amplitude for $\varphi=0$ or $\pi$ and the largest one for $\varphi=\pm \pi/2$.

\section{EFFECTIVE FILED THEORY } \label{field theory}
In this section, we introduce a low-energy effective action to describe all three phases, i.e., NSF, $D_4$SF and MI phases. Based on this action, we will show phase diagram calculated by mean-field theory and a critical exponent calculated by momentum shell RG theory.

In order to construct the effective action, two important factors from microscopic analysis above must be considered. First,  kinetic energy has the quartic form of $k_x^4+k_y^4+a ( k_x^2+k_y^2 )$ at small momentum. Second, momentum-dependent interaction has the quadratic form of $\alpha+\beta (k_x^2+k_y^2) $ at small momentum. These two factors also agree with $D_4$ symmetry. The low-energy effective action of the $d$-dimensional lattice shaken along the $x$ and $y$ directions can be written as
\begin{equation}
\mathcal{S}(\Phi, \Phi^{*})= \int_0^{1/T} \mathrm{d} \tau \int \mathrm{d}^d \bm{r}  \{K_1 \Phi^{*} \partial_{\tau} \Phi+K_2 |\partial_{\tau} \Phi|^2+\mathcal{E}(\Phi,\Phi^{*}) \},     \label{action}
\end{equation}
where 
\begin{align}
\mathcal{E}(\Phi,\Phi^{*})&=|\partial_x^2 \Phi|^2+|\partial_y^2 \Phi |^2 +a |\nabla \Phi |^2+\mathcal{T}+r | \Phi |^2  \nonumber \\
& \quad +\alpha | \Phi |^4+\beta | \Phi \nabla \Phi |^2, 
\end{align}
$T$ is temperature, $\Phi$ is superfluid order parameter, $\nabla=(\partial_x,\partial_y)$, $\mathcal{T}=0$ for $d=2$ and $\mathcal{T}=|\partial_z \Phi |^2$ for $d=3$.  The signs of parameters $a$ and $r$ can be inverted by tuning shaking amplitude $f$ and interaction strength $g$, respectively. Parameter $\alpha$ is considered to be positive for repulsive interactions. And parameter $\beta$ can be either positive in the deep lattice limit, or negative in the shallow lattice limit. We will take $\beta>0$ for simplicity. 

\begin{figure}[tbp]
\centering
\includegraphics[height=2in,width=2in]
{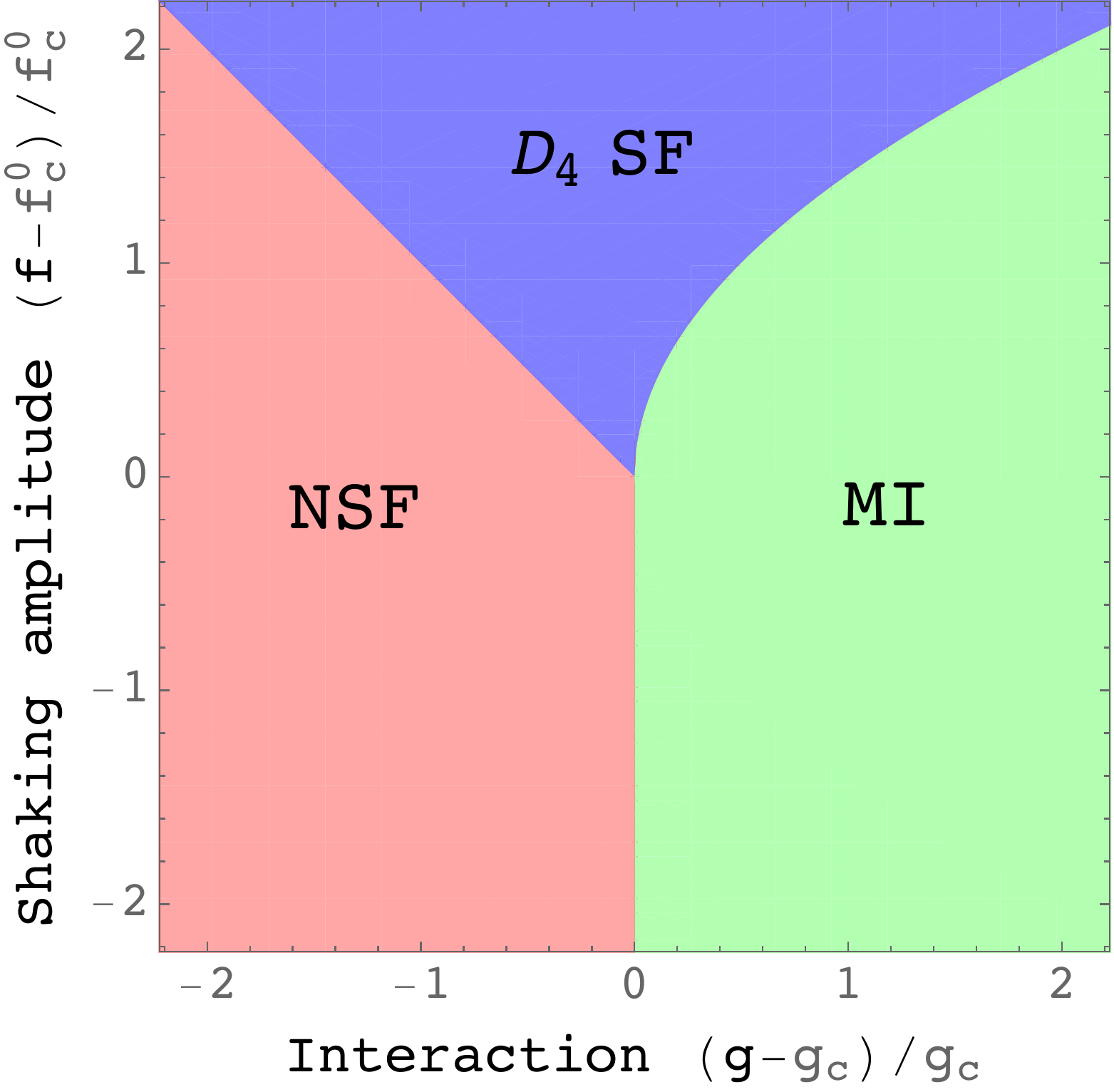}
\caption{Mean-field phase diagram. $g_c$ is the critical interaction strength for the NSF-MI transition. $f_c^0$ is the critical shaking amplitude calculated by minimizing the single-particle quasienergy dispersion. }
\label{phase diagram}
\end{figure}

Assuming $\Phi=|\Phi| e^{i \bm{k} \cdot \bm{r}}$, $\mathcal{E}$ can be rewritten as
\begin{equation}
\mathcal{E}( | \Phi |, \bm{ k } )=(k_x^4+k_y^4+a \bm{k}^2+r) |\Phi |^2+(\alpha+\beta \bm{k}^2) |\Phi |^4,
\end{equation}
where $\bm{k}=(k_x,k_y, \cdots)$ and $\bm{r}=(x,y, \cdots)$ are $d$-dimensional vectors. By minimizing $\mathcal{E}$ with respect to $|\Phi |$ and $\bm{k}$, one obtains three different phases: (1) the MI phase with $ \Phi = 0 $; (2) the NSF phase with $ \Phi \neq 0 $ and $ \bm{k}=0 $; (3) the $D_4$SF phase with $ \Phi \neq 0, k_x \neq 0, k_y \neq 0 $ and $ \bm{k}^2-k_x^2-k_y^2=0$. Phase boundaries are also obtained: (1) $ r=0 $ and $ a > 0 $ separating the NSF and MI phase; (2) $ r=2 \alpha a/\beta < 0 $ separating the NSF and $D_4$SF phase; (3) $r=a^2/2 $ and $ a < 0 $ separating the $D_4$SF and MI phase, which is different from the 1D case \cite{ising}. There is a mean-field tricritical point $(a,r)=(0,0)$. In the vicinity of this tricritial point, $a$ and $r$ are proportional to $ f - f_c^0 $ and $ g - g_c $, respectively. The phase diagram in $f$- and $g$-terms is shown in Fig. (\ref{phase diagram}).

Next we will study the critical correlation length exponent within momentum shell RG approach.

At zero temperature and in $(d+1)$ dimensional momentum and frequency space, the action in Eq.(\ref{action}) can be rewritten as
\begin{align}
S(\Phi,\Phi^{*})&=\int \frac{\mathrm{d}^d \bm{k}}{(2 \pi)^{\mathrm{d}}}  \frac{\mathrm{d} \omega}{2 \pi}
\Phi^{*}(\bm{k},\omega) \{-i K_1 \omega+K_2 \omega^2     \nonumber \\
& \quad +k_x^4+k_y^4 +a (k_x^2+k_y^2)+\mathcal{T}_k+r  \}  \Phi(\bm{k},\omega)  \nonumber \\
& \quad +\int_{\bm{k} \omega}^{\Lambda} \{ \alpha+\beta (k_{2x} k_{4x}+k_{2y} k_{4y}) \} \Phi^{*}(\bm{k_1},\omega_1)   \nonumber \\
& \quad \Phi^{*}(\bm{k_2},\omega_2) \Phi(\bm{k_3},\omega_3) \Phi(\bm{k_4},\omega_4),    \label{origin action}
\end{align}
where $\mathcal{T}_k=0$ for $d=2$, $\mathcal{T}_k=k_z^2$ for $d=3$,  $\int_{\bm{k} \omega}^{\Lambda}=\int^{\Lambda}  \frac{\prod_{i=1}^4 \mathrm{d}^d \bm{k_i} \mathrm{d} \omega_i }{(2 \pi)^{3\mathrm{d}+3}}  \delta(\bm{k_1}+\bm{k_2}-\bm{k_3}-\bm{k_4}) \delta(\omega_1+\omega_2-\omega_3-\omega_4)$ and $\Lambda$ denotes high momentum cut-off. 

\begin{figure}[tbp]
\centering
\includegraphics[height=1.35in,width=3.2in]
{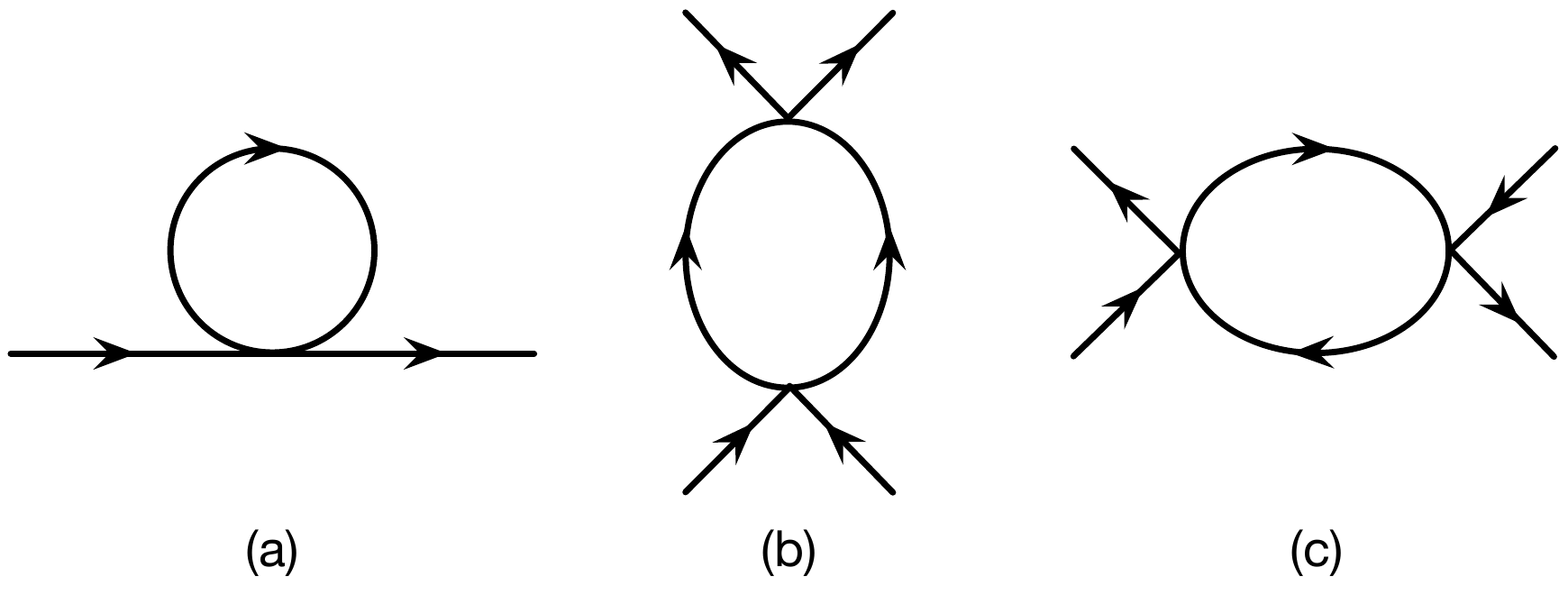}
\caption{ The one-loop Feynman graphs. Graph (a) contributes to renormalizing  parameters $a$ and $r$. Graph (b) and (c) contribute to renormalizing  parameters $\alpha$ and $\beta$.  }
\label{feynman}
\end{figure}

The one-loop Feynman graphs for renormalizing the parameters in Eq.(\ref{origin action}) are shown in Fig. (\ref{feynman}). The study of the critical exponent is divided into two cases.

Case $\mathbf{A}$. Without particle-hole symmetry. $K_1 \neq 0$, so the $K_2$-term becomes irrelevant. In this case, scaling dimensions of the parameters read
\begin{align}
& [k_x]=[k_y]=\frac{1}{2}, [k_z]=1, [a]=1, [r]=2, [\omega]=2,   \nonumber \\
& [\alpha]=3-d, [\beta]=2-d,  [\Phi]=-\frac{d+3}{2}.
\end{align}
The upper critical dimension is 3, which is $1/2$ larger than that in the 1D shaken lattice \cite{ising} due to the extra shaking direction. For $d=2$, the $\alpha$-term is relevant and the $\beta$-term is marginal. It's different from the irrelevant $\beta$-term in the 1D case \cite{ising}. So we need to consider corrections from the $\beta$-term. For graphs in Fig.\ref{feynman} (b,c), we need to expand them in powers of external momenta. The one-loop RG flow equations read
\begin{align}
\frac{ \mathrm{d} a }{ \mathrm{d} l } & =a ,  \\
\frac{ \mathrm{d} r }{ \mathrm{d} l } & =2r ,  \\
\frac{ \mathrm{d} \alpha }{ \mathrm{d} l } & = \epsilon \alpha-\frac{ \alpha^2 }{ 1+r } I_2(a) ,
\\
\frac{ \mathrm{d} \beta }{ \mathrm{d} l } & = (\epsilon-1) \beta - \frac{ \alpha \beta }{ 1+r } I_2(a) + \frac{1}{ (1+r)^2 }  
\left[ \alpha^2 J_2(a) \right.  \nonumber
\\
&  \quad  \left. - \alpha \beta L_2(a) \right] - \frac{ \alpha^2 }{ (1+r)^3 } M_2(a),
\end{align}
where
\begin{widetext}
\begin{align}
I_2(a)&=\int_0^{2 \pi} \frac{ \mathrm{d} \phi }{ (2 \pi)^2 }  
 \frac{ -a+\sqrt{ a^2+4 ( \cos^4 \phi + \sin^4 \phi)} }{ 4 ( \cos^4 \phi + \sin^4 \phi)},
\\
J_2(a)&=\int_0^{2 \pi} \frac{ \mathrm{d} \phi }{ (2 \pi)^2 } 
 \frac{ -a+\sqrt{ a^2+4 ( \cos^4 \phi + \sin^4 \phi)} }{ 2 ( \cos^4 \phi + \sin^4 \phi ) } 
 \left[ a + 3 \cos^2 \phi  \frac{ -a+\sqrt{ a^2+4 ( \cos^4 \phi + \sin^4 \phi)} }{  \cos^4 \phi + \sin^4 \phi }   \right],
\\
L_2(a)&=\int_0^{2 \pi} \frac{ \mathrm{d} \phi }{ (2 \pi)^2 } 
\left[ \frac{ -a+\sqrt{ a^2+4 ( \cos^4 \phi + \sin^4 \phi)} }{ 2 ( \cos^4 \phi + \sin^4 \phi ) } \right]^2
\left[ a + \cos^2 \phi  \frac{ -a+\sqrt{ a^2+4 ( \cos^4 \phi + \sin^4 \phi)} }{  \cos^4 \phi + \sin^4 \phi }   \right]  \cos^2 \phi ,
\\
M_2(a) & = \int_0^{2 \pi} \frac{ \mathrm{d} \phi }{ (2 \pi)^2 } 
\left[ \frac{ -a+\sqrt{ a^2+4 ( \cos^4 \phi + \sin^4 \phi)} }{  \cos^4 \phi + \sin^4 \phi } \right]^2
\left[ a + \cos^2 \phi  \frac{ -a+\sqrt{ a^2+4 ( \cos^4 \phi + \sin^4 \phi)} }{  \cos^4 \phi + \sin^4 \phi }   \right]^2   \frac{ \cos^2 \phi }{2}
\end{align}
\end{widetext}
and $\epsilon=3-d$.

\begin{figure}[!tbp]
\centering
\includegraphics[height=1.35in,width=1.6in]{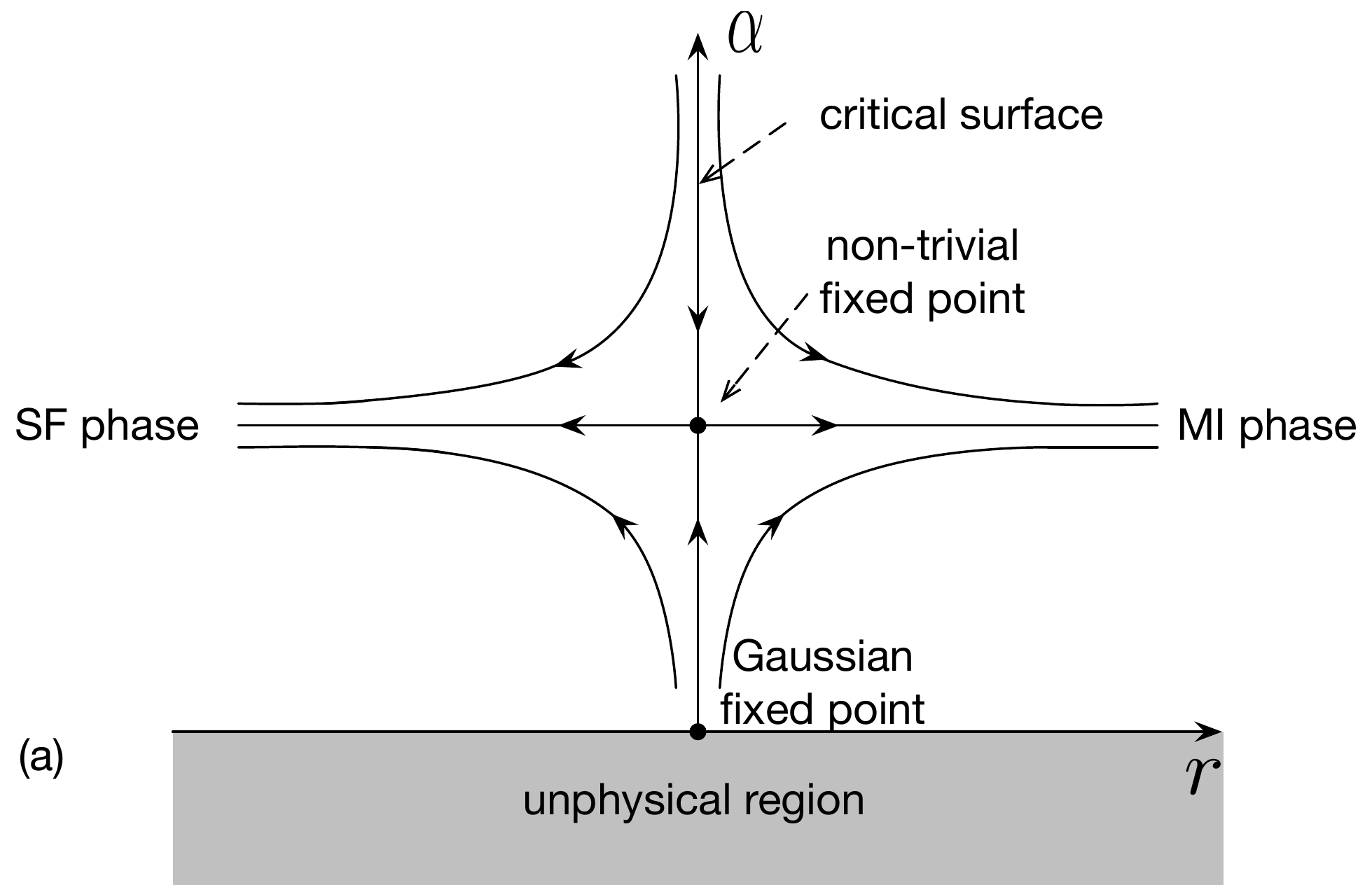}
\includegraphics[height=1.35in,width=1.6in]{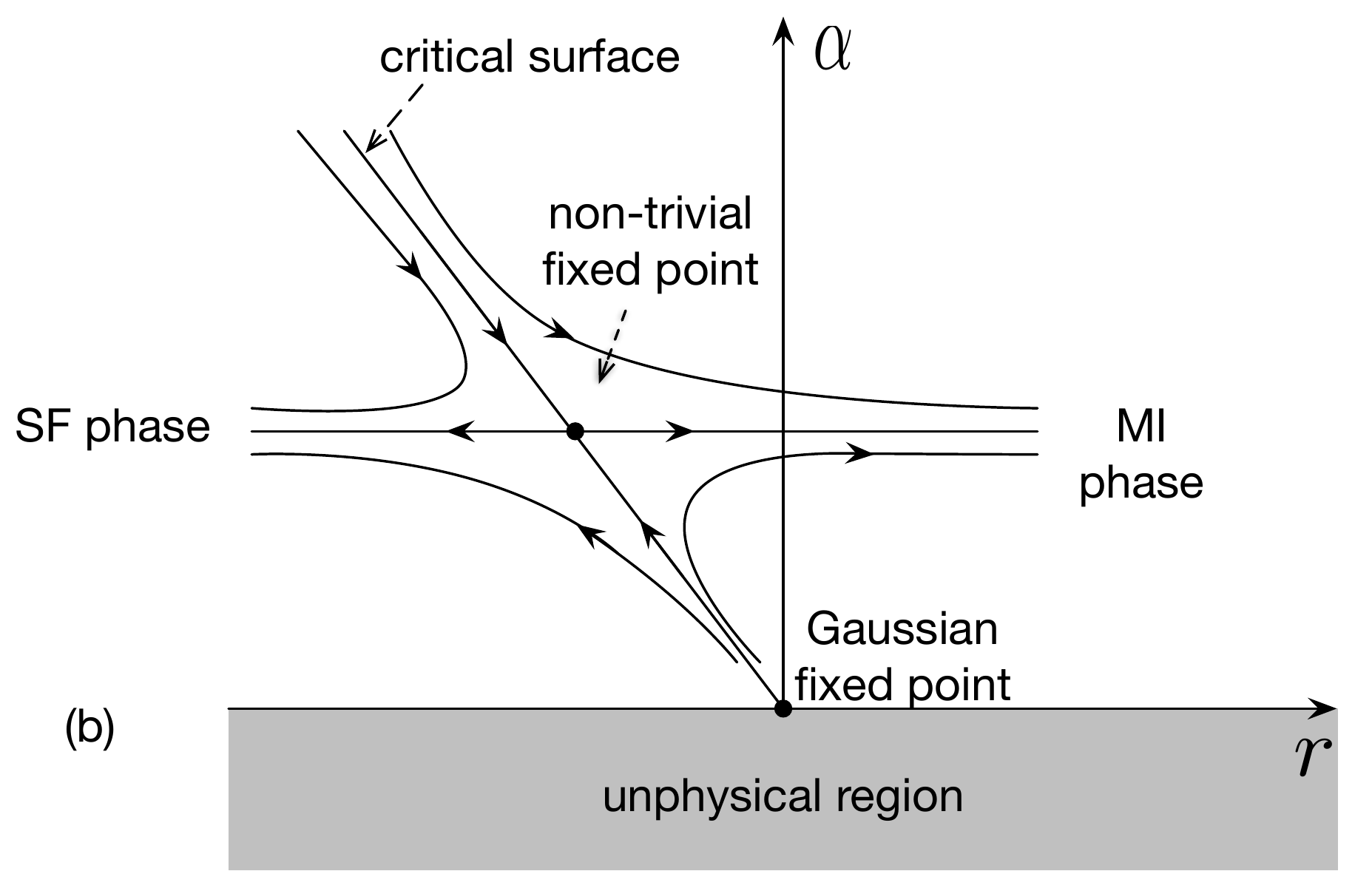}
\caption{RG flow diagrams of (a) case \textbf{A} and (b) case \textbf{B} on the $r-\alpha$ plane. } 
\label{flow}
\end{figure}

The nontrivial fixed point lies at $ ( r^{\ast}, a^{\ast}, \alpha^{\ast} ,\beta^{\ast})=( 0, 0, \frac{ 1 }{ I_2(0) } \epsilon,  \frac{J_2(0) - M_2(0) }{ I_2(0)^2 } \epsilon^2 )$. Defining new variables $ \delta r = r - r^{\ast}, \delta a = a - a^{\ast}, \delta \alpha=\alpha-\alpha^{\ast} ,  \delta \beta=\beta-\beta^{\ast}$, the linearized flow equations are given by
\begin{equation} \label{case a}
\frac{ \mathrm{d} }{ \mathrm{d} l }
\left( \begin{array}{c}
\delta r \\ 
\delta a  \\
\delta \alpha \\
\delta \beta
\end{array} \right)
= 
\left(
\begin{array}{cccc}
2 & 0 & 0 & 0 \\
0 & 1 & 0  & 0 \\
0 & 0 & -\epsilon &0 \\
0 & 0& 2 \frac{ J_2(0) - M_2(0) }{ I_2(0) } \epsilon & -1- \frac{ L_2(0) }{ I_2(0) } \epsilon
\end{array}
\right)
\left( \begin{array}{c}
\delta r \\ 
\delta a  \\
\delta \alpha \\
\delta \beta
\end{array} \right).
\end{equation}
Eigenvalues of the matrix in Eq.(\ref{case a}) are $2, 1, - \epsilon, -1- \epsilon  L_2(0) / I_2(0)$. Then the scaling dimension of parameter $r$ at the nontrivial fixed point is $y_r=2$. The correlation length exponent of the superfluid transition can be calculated as $\nu=1/y_r=1/2$. It is the same as the mean-field value in the usual Bose gas \cite{exponent1,exponent2} and the value in the 1D case with $K_1 \neq 0$ \cite{ising} because of no one-loop corrections on $r$ from interaction as shown in the flow diagram in Fig.\ref{flow}(a).

Case $\mathbf{B}$. With particle-hole symmetry. $K_1= 0$. In this case, scaling dimensions of the parameters read
\begin{align}
& [k_x]=[k_y]=\frac{1}{2}, [k_z]=1, [a]=1, [r]=2, [\omega]=1,   \nonumber \\
& [\alpha]=4-d, [\beta]=3-d,  [\Phi]=-\frac{d+2}{2}.
\end{align}
The upper critical dimension is 4. For $d=3$, the one-loop RG equations read
\begin{align}
\frac{ \mathrm{d} a }{ \mathrm{d} l } & =a+ \frac{\beta}{2 \sqrt{1+r} } I_3(a), 
 \\
\frac{ \mathrm{d} r }{ \mathrm{d} l } & =2r+\frac{ 1 }{ \sqrt{ 1+r }  } \left[ 2 \alpha I_3(a) + \frac{ \beta }{2} J_3(a) \right] ,  
\\
\frac{ \mathrm{d} \alpha }{ \mathrm{d} l } & = \epsilon \alpha-\frac{ 1}{ (1+r)^{ \frac{3}{2} } } \left[ \frac{5}{2} \alpha^2  I_3(a) + \alpha \beta J_3(a)+ \beta^2 L_3(a) \right] ,  
\end{align}
\begin{widetext}
\begin{align}
\frac{ \mathrm{d} \beta }{ \mathrm{d} l }  & = (\epsilon-1) \beta -\frac{1}{2 (1+r)^{\frac{3}{2} }} \left[ \alpha \beta I_3(a) + \frac{ \beta^2 }{2}  J_3(a) \right]  
+ \frac{1}{ (1+r)^{ \frac{5}{2}} } \left\{ \frac{9}{2} \alpha^2 \left[ a I_3(a) + 3 J_3(a) \right]  +  \alpha \beta M_3(a) + \beta^2 N_3(a) \right\}   \nonumber
\\
& \quad  + \frac{1}{ (1+r)^{ \frac{7}{2} } } \left[ \alpha^2 P_3(a) + \alpha \beta R_3(a) + \beta^2 S_3(a) \right],
\end{align}
where 
\begin{align}
I_3(a) &= \frac{1}{(2 \pi)^3 }\int_0^{\pi} \mathrm{d} \theta \int_0^{2 \pi}  \mathrm{d} \phi   \sqrt{ \frac{ \sin^2 \theta }{4}+ \cos^2 \theta}  
  \left[ \frac{ -( a \sin^2 \theta+ \cos^2 \theta )+\sqrt{ (a \sin^2 \theta+ \cos^2 \theta)^2+4 \sin^4 \theta (\sin^4 \phi +\cos^4 \phi)} }{ 2 \sin^4 \theta ( \sin^4 \phi +\cos^4 \phi) } \right]^{\frac{3}{2}} \nonumber
  \\
  & \quad \sin \theta,
 \\
 J_3(a) &= \frac{1}{(2 \pi)^3 } \int_0^{\pi} \mathrm{d} \theta \int_0^{2 \pi}  \mathrm{d} \phi  \sqrt{ \frac{ \sin^2 \theta }{4}+ \cos^2 \theta}  
  \left[ \frac{ -( a \sin^2 \theta+ \cos^2 \theta )+\sqrt{ (a \sin^2 \theta+ \cos^2 \theta)^2+4 \sin^4 \theta (\sin^4 \phi +\cos^4 \phi)} }{ 2 \sin^4 \theta ( \sin^4 \phi +\cos^4 \phi) } \right]^{\frac{5}{2}} \nonumber
  \\
  & \quad \sin^3 \theta ,
    \\
 L_3(a) & = \frac{1}{ 8 (2 \pi)^3 } \int_0^{\pi} \mathrm{d} \theta \int_0^{2 \pi}  \mathrm{d} \phi  \sqrt{ \frac{ \sin^2 \theta }{4}+ \cos^2 \theta}  
  \left[ \frac{ -( a \sin^2 \theta+ \cos^2 \theta )+\sqrt{ (a \sin^2 \theta+ \cos^2 \theta)^2+4 \sin^4 \theta (\sin^4 \phi +\cos^4 \phi)} }{ 2 \sin^4 \theta ( \sin^4 \phi +\cos^4 \phi) } \right]^{\frac{7}{2}} \nonumber
  \\
  & \quad \sin^5 \theta ,
 \\
M_3(a) &= \frac{3}{ 2 (2 \pi)^3 }  \int_0^{\pi} \mathrm{d} \theta \int_0^{2 \pi}  \mathrm{d} \phi \sqrt{ \frac{ \sin^2 \theta }{4}+ \cos^2 \theta}  
  \left[ \frac{ -( a \sin^2 \theta+ \cos^2 \theta )+\sqrt{ (a \sin^2 \theta+ \cos^2 \theta)^2+4 \sin^4 \theta (\sin^4 \phi +\cos^4 \phi)} }{ 2 \sin^4 \theta ( \sin^4 \phi +\cos^4 \phi) } \right]^{\frac{5}{2}}     \nonumber
  \\
  & \quad \sin^3 \theta 
 \left[ a ( 1+\cos^2 \phi ) + \cos^2 \phi ( 7 + \cos 2 \phi )  \frac{ -( a \sin^2 \theta+ \cos^2 \theta )+\sqrt{ (a \sin^2 \theta+ \cos^2 \theta)^2+4 \sin^4 \theta (\sin^4 \phi +\cos^4 \phi)} }{ 2 \sin^2 \theta ( \sin^4 \phi +\cos^4 \phi) }  \right] ,
 \\
 N_3(a) &= \frac{3}{ 16 (2 \pi)^3 }\int_0^{\pi} \mathrm{d} \theta \int_0^{2 \pi}  \mathrm{d} \phi   \sqrt{ \frac{ \sin^2 \theta }{4}+ \cos^2 \theta}  
  \left[ \frac{ -( a \sin^2 \theta+ \cos^2 \theta )+\sqrt{ (a \sin^2 \theta+ \cos^2 \theta)^2+4 \sin^4 \theta (\sin^4 \phi +\cos^4 \phi)} }{ 2 \sin^4 \theta ( \sin^4 \phi +\cos^4 \phi) } \right]^{\frac{7}{2}} \nonumber
  \\
  & \quad \sin^5 \theta 
  \left[ a ( 3 + 2 \cos 2 \phi ) + \cos^2 \phi ( 5 + 2 \cos 2 \phi) \frac{ -( a \sin^2 \theta+ \cos^2 \theta )+\sqrt{ (a \sin^2 \theta+ \cos^2 \theta)^2+4 \sin^4 \theta (\sin^4 \phi +\cos^4 \phi)} }{ \sin^2 \theta ( \sin^4 \phi +\cos^4 \phi) }  \right],
 \\
 P_3(a) &=  \frac{5}{(2 \pi)^3}   \int_0^{\pi} \mathrm{d} \theta \int_0^{2 \pi}  \mathrm{d} \phi  \sqrt{ \frac{ \sin^2 \theta }{4}+ \cos^2 \theta}  
  \left[ \frac{ -( a \sin^2 \theta+ \cos^2 \theta )+\sqrt{ (a \sin^2 \theta+ \cos^2 \theta)^2+4 \sin^4 \theta (\sin^4 \phi +\cos^4 \phi)} }{ 2 \sin^4 \theta ( \sin^4 \phi +\cos^4 \phi) } \right]^{\frac{5}{2}} \nonumber 
\\
& \quad \sin^3 \theta \cos^2 \phi
\left[ a + \cos^2 \phi \frac{ -( a \sin^2 \theta+ \cos^2 \theta )+\sqrt{ (a \sin^2 \theta+ \cos^2 \theta)^2+4 \sin^4 \theta (\sin^4 \phi +\cos^4 \phi)} }{ \sin^2 \theta ( \sin^4 \phi +\cos^4 \phi) }  \right],
\\
R_3(a) & =  \frac{5}{(2 \pi)^3}   \int_0^{\pi} \mathrm{d} \theta \int_0^{2 \pi}  \mathrm{d} \phi  \sqrt{ \frac{ \sin^2 \theta }{4}+ \cos^2 \theta}  
  \left[ \frac{ -( a \sin^2 \theta+ \cos^2 \theta )+\sqrt{ (a \sin^2 \theta+ \cos^2 \theta)^2+4 \sin^4 \theta (\sin^4 \phi +\cos^4 \phi)} }{ 2 \sin^4 \theta ( \sin^4 \phi +\cos^4 \phi) } \right]^{\frac{7}{2}} \nonumber 
\\
& \quad \sin^5 \theta \cos^2 \phi
\left[ a + \cos^2 \phi  \frac{ -( a \sin^2 \theta+ \cos^2 \theta )+\sqrt{ (a \sin^2 \theta+ \cos^2 \theta)^2+4 \sin^4 \theta (\sin^4 \phi +\cos^4 \phi)} }{ \sin^2 \theta ( \sin^4 \phi +\cos^4 \phi) } \right]^2 ,
\\
S_3(a) & = \frac{5}{ 8 (2 \pi)^3}   \int_0^{\pi} \mathrm{d} \theta \int_0^{2 \pi}  \mathrm{d} \phi  \sqrt{ \frac{ \sin^2 \theta }{4}+ \cos^2 \theta}  
  \left[ \frac{ -( a \sin^2 \theta+ \cos^2 \theta )+\sqrt{ (a \sin^2 \theta+ \cos^2 \theta)^2+4 \sin^4 \theta (\sin^4 \phi +\cos^4 \phi)} }{ 2 \sin^4 \theta ( \sin^4 \phi +\cos^4 \phi) } \right]^{\frac{9}{2}} \nonumber 
\\
& \quad \sin^7 \theta \cos^2 \phi
\left[ a + \cos^2 \phi  \frac{ -( a \sin^2 \theta+ \cos^2 \theta )+\sqrt{ (a \sin^2 \theta+ \cos^2 \theta)^2+4 \sin^4 \theta (\sin^4 \phi +\cos^4 \phi)} }{ \sin^2 \theta ( \sin^4 \phi +\cos^4 \phi) } \right]^2 
\end{align}
\end{widetext}
and $\epsilon=4-d=1$. The nontrivial fixed point lies at  
\begin{align}
 ( r^{\ast}, a^{\ast}, \alpha^{\ast}, \beta^{\ast} ) & = \left( - \frac{ 2 \epsilon }{ 5 }, - \frac{  27 J_3(0) + 2 P_3(0) }{ 25 I_3(0) } \epsilon^2  ,  \frac{ 2 \epsilon }{ 5 I_3(0) }, \right.
 \nonumber 
 \\
& \qquad \left.  \frac{ 2  \left[ 27 J_3(0) + 2 P_3(0) \right] }{ 25 I_3^2(0) } \epsilon^2 \right).
 \end{align}
Defining $ \delta r=r-r^{\ast},  \delta a=a-a^{\ast},  \delta \alpha=\alpha-\alpha^{\ast} $ and $ \delta \beta = \beta - \beta^{\ast} $ gives the linearized equations
\begin{widetext}
\begin{align}
&\frac{ \mathrm{d} }{ \mathrm{d} l }
\left( \begin{array}{c}
\delta r \\ 
\delta a \\
\delta \alpha  \\
\delta \beta 
\end{array} \right) 
= 
\left(
\begin{array}{cccc}
2-\frac{ 2 }{5} \epsilon  &  \frac{ 4 }{ 5 I_3(0) } \frac{ \partial I_3(a) }{ \partial a } \lvert_{a=0}  \epsilon  &  2 I_3(0) ( 1 + \frac{1}{5} \epsilon )   &  \frac{ J_3(0) }{2} ( 1 + \frac{1}{5} \epsilon )      \\
0  &  1  &  0  &  \frac{ I_3(0) }{2} ( 1 + \frac{1}{5} \epsilon )     \\
0  &  0  &  -\epsilon  &  - \frac{ 2 J_3(0) }{ 5 I_3(0) } \epsilon     \\
0  &  0  &  \left[ 27 J_3(0)  + 2 P_3(0) \right]  \frac{ 2  }{ 5 I_3(0) } \epsilon &  -1 +  \frac{ 2  }{5} \left[ 2 +  \frac{ M_3(0)  + R_3(0) }{ I_3(0) } \right] \epsilon
\end{array}
\right) 
 \left( \begin{array}{c}
\delta r \\ 
\delta a \\
\delta \alpha  \\
\delta \beta 
\end{array} \right).    \label{flow equation}
\end{align}
\end{widetext}
The eigenvalues of the matrix in Eq.(\ref{flow equation}) are $2-2 \epsilon/5, 1, -\epsilon$ and $ - 1 + 4 \epsilon /5 + 2 \epsilon ( M_3(0) + R_3(0) ) / 5 I_3(0)$. Here it is the important difference from Case $\mathbf{A}$ that the $r$-term gets corrections from interaction as shown in flow diagram in Fig. (\ref{flow})(b). The correlation length exponent of the superfluid transition is $\nu=1/(2-2 \epsilon /5)=1/2+\epsilon /10=3/5$.  It is different from bosons with quartic dispersion in only one direction with particle-hole symmetry \cite{ising} due to different upper critical dimensions, while it is the same as conventional bosons with quadratic dispersion with particle-hole symmetry in three dimension, and thus, belongs to the $O(2)$ rotor model class, up to $\epsilon$ order \cite{rotor}. 

Systems with parameters $r$ and $\alpha$ that lie in the right region of the critical surface in Fig. (\ref{flow}) will eventually flow towards the MI phase, whereas systems with $r$ and $\alpha$ in the left region of the critical surface will flow towards the NSF phase for the initial $a>a^{\ast}$ or the $D_4$SF phase for the initial $a<a^{\ast}$.

\section{The fate of the BEC in general shaken lattice systems} \label{non condensate}
In sections above, we suppose a BEC exists in the system. However, it is known that low-energy density of states can be increased by SO coupling \cite{hui hu, gordon baym, qi zhou, qi zhou2, bose liquid12} or lattice shaking \cite{bose liquid12}. Fluctuations may destroy off-diagonal long-range order (ODLRO). In this section we will first study existence of ODLRO in the 2D cases, and then in general cases. 

Let us first assume bosons condensing in the NSF phase and write $\Phi= \sqrt{\rho_0+\delta \rho} e^{ i \theta }$, where $\rho_0= - r /2 \alpha $ is superfluid density, $\delta \rho$ is density fluctuation and $ \theta $ is phase fluctuation. Then we substitute the $\Phi$ field into Eq.(\ref{action}) and expand the action in Eq.(\ref{action}) to quadratic order in $\delta \rho$ and $\theta$. By integrating out $\delta \rho$ field, the low-energy effective action for $\theta$ field is given by
\begin{align}
\mathcal{S}_{eff} ( \theta ) & = \rho_0 \int \mathrm{d}^d r \mathrm{d} \tau \Big\{ \tilde{K} ( \partial_{\tau} \theta )^2 + \partial_x^4 \theta + \partial_y^4 \theta + \tilde{a} ( \nabla \theta )^2  \nonumber \\
& \quad + \mathcal{T} \Big\},       
\end{align}
where $\tilde{K}=  \frac{K_1^2}{ 4 \alpha \rho_0 }+ K_2 , \tilde{a} = a+\beta \rho_0 $. 

In Gaussian approximation, the correlation function can be written as
\begin{equation}
\langle \Phi^{\ast}( \bm{r}) \Phi ( 0 ) \rangle 
= \rho_0  e^{ -\frac{1}{2} \langle ( \theta(\bm{r})-\theta(0) )^2 \rangle }   .               \label{correlation}
\end{equation}

At finite temperature, Eq.(\ref{correlation}) can be written as
\begin{align}
\langle \Phi^{\ast}( \bm{r}) \Phi ( 0 ) \rangle
&= \rho_0 \mathrm{exp} \bigg\{ - T \int \frac{ \mathrm{d}^d \bm{k} }{ (2 \pi)^d } ( 1-e^{i \bm{kr} } )   \nonumber \\
& \quad  \left( k_x^4+k_y^4+ \tilde{a} (k_x^2+ k_y^2)+\mathcal{T}_k \right)^{-1} \bigg\}.   \label{finite T}
\end{align} 

For $d=2$, in the large separation limit, the integral in Eq.(\ref{finite T}) can be approximated by
\begin{align}
  \int_0^{2 \pi} \mathrm{d} \theta \int_0^{\Lambda}   \frac{\mathrm{d} k}{ (2 \pi)^2 }  \frac{1}{ k^3 ( \cos^4 \theta+\sin^4 \theta )+ \tilde{a} k }.  \label{integral1}
\end{align} 
As $k \rightarrow 0$, the integrand in Eq.(\ref{integral1}) behaves as $1/k$ for $\tilde{a} > 0$ and $1/k^3$ for $\tilde{a}=0$. So the integral is divergent, the correlation function in Eq.(\ref{finite T}) approaches zero at large separation, and ODLRO is absent at any finite temperature.

For $d=3$ and $\tilde{a} > 0$, in the large separation limit, the integral in Eq.(\ref{finite T}) can be approximated by
\begin{align}
\int \frac{ \mathrm{d}^3 \bm{k} }{ (2 \pi)^3 }
\frac{1}{  k_x^4+k_y^4+ \tilde{a} (k_x^2+ k_y^2)+ k_z^2 } .
\end{align}
After integrating over $k_z$, the integral above is given by
\begin{align}
 \int_0^{2 \pi} \mathrm{d} \theta  \int_0^{\Lambda}   \frac{\mathrm{d} k}{ 8 \pi^2 }  \frac{1}{ \sqrt{ k^2 ( \cos^4 \theta+\sin^4 \theta )+ \tilde{a} } }.  \label{integral2}
\end{align} 
As $k \rightarrow 0$, the integrand in Eq.(\ref{integral2}) behaves like  $1$. So the integral is finite and ODLRO exists.

For $d=3$ and $\tilde{a} = 0$,  the integral in the exponent in Eq.(\ref{finite T}) reads
\begin{align}
 \int \frac{ \mathrm{d}^3 \bm{k} }{ (2 \pi)^3 } \frac{1-e^{i \bm{k} \bm{r} } }  
 { k_x^4+k_y^4+k_z^2 } .   
\end{align}
The critical point $ \tilde{a}=0 $ means critical shaking amplitude $f_c$ in Sec. \ref{fc}, i.e., phase boundary separating NSF and $D_4$SF phases in the mean field level in Sec. \ref{field theory}. At large separation, the correlation function is given by
\begin{align}
\langle \Phi^{\ast}(x) \Phi ( 0 ) \rangle & \sim  |x|^{ - 2 \eta T },
\\
\langle \Phi^{\ast}(z) \Phi ( 0 ) \rangle & \sim  | z |^{ - \eta T } .
\end{align}
where $\eta =  \Gamma( 5/4 )^2 / \pi^{ 5/2 } $ and $\Gamma(z)$ denotes Euler gamma function. So there exists BEC only at zero temperature and noncondensed Bose liquid at finite temperature. This algebraically ordered Bose liquid is anisotropic. Since fluctuation is enhanced by shaking, the correlation function decays faster along the shaking directions. 

At zero temperature, Eq.(\ref{correlation}) can be written as
\begin{align} \label{correlation_zero_tem}
\langle \Phi^{\ast}( \bm{r}) \Phi ( 0 ) \rangle
&= \rho_0  \mathrm{exp} \bigg\{ - \int \frac{ \mathrm{d}^d \bm{k} \mathrm{d} \omega }{ (2 \pi)^{ d+1 } } ( 1-e^{i \bm{kr} } )  \nonumber \\
& \quad \left( \tilde{K} \omega^2+ k_x^4+k_y^4+ \tilde{a} (k_x^2+ k_y^2)+\mathcal{T}_k \right)^{-1} \bigg\}.
\end{align} 
The zero temperature results are equivalent to adding a unshaken direction to the corresponding finite temperature case. The correlation function in Eq.(\ref{correlation_zero_tem}) remains finite at large separation for $\tilde{a} > 0$ and $d=2$ and $3$. At the critical point $\tilde{a}=0$ and in the large separation limit, the vanishing correlation function is found for $d=2$, which is consistent with results in SO-coupled BEC with similar dispersion \cite{bose liquid11, bose liquid12}, whereas the finite correlation function is found for $d=3$. 

From calculations above, we know phase fluctuation destroys ODLRO at any finite temperature for $d=2$ and $ \tilde{a}>0 $. The effect is even stronger at the critical point due to pure quartic dispersion. For $d=2$ and $ \tilde{a}=0 $, ODLRO does not exists even at zero temperature \cite{bose liquid11, bose liquid12}. For a system with $d=3$,  there exists ODLRO when $ \tilde{a}>0 $ and quasi-long-range order at finite temperature when $ \tilde{a}=0 $. Therefore the BEC can be changed into a non-condensed Bose liquid by tuning the shaking amplitude $f$ approaching the critical value $f_c$. 

Raman-induced SO coupling and lattice shaking have generated quartic dispersion \cite{bose liquid12}. And higher-order terms may be generated in the future. Next we will study the feasibility of changing the BEC into a non-condensed Bose liquid in a system with a general dispersion. 
 
\begin{table}[!htb]
\caption{\label{order table} Existence of ODLRO for $ \tilde{a}=0 $ at finite temperatures.}
\begin{ruledtabular}
\begin{tabular}{ c  c  c  c  c } 
    &   
\multirow{2}*{ $\mathcal{T}_k$ }    &  
$k_x^n+\mathcal{T}_k$    &    
$k_x^n+k_y^m+\mathcal{T}_k$  & 
$k_x^n+k_y^m+k_z^l$        \\

  & 
  &    
$( n \geq 4 )$    &    
$( m \geq n \geq 4 )$    &     
$(l \geq m \geq n \geq 4)$   \\ \hline

d=2   &  
N  &  
 N  &  
N  &   
---    \\ \hline

d=3    &  
O  &   
O  &    
N &  
N  \\
\end{tabular}          \label{finite temperature}    
\end{ruledtabular}

\caption{\label{order table} Existence of ODLRO for $ \tilde{a}=0 $ at zero temperature.}
\begin{ruledtabular}
\begin{tabular}{ c c c  c  c } 
    &   
\multirow{2}*{ $\mathcal{T}_k$ }    &   
$k_x^n+\mathcal{T}_k$    &    
$k_x^n+k_y^m+\mathcal{T}_k$   & 
$k_x^n+k_y^m+k_z^l$        \\

 & 
 &
$( n \geq 4 )$    &    
$( m \geq n \geq 4 )$    &     
$(l \geq m \geq n \geq 4)$   \\ \hline

d$=2$   &  
O &   
O &  
N &   
---   \\ \hline

\multirow{3}*{d$=3$}    &    
\multirow{3}*{O}     &    
\multirow{3}*{O}     &     
\multirow{3}*{O}   &   
O: $n=m=4$        \\ 

  &  &  &  &   or  $n=4, m=6, l \leq 10$;     \\

  &  &  &  &   N: otherwise  \\ 
\end{tabular}          \label{zero temperature}
\end{ruledtabular}
\end{table}     

The existence of ODLRO is obtained by checking if the correlation function in Eq.(\ref{correlation}) is finite in large separation limit. And the results at the critical point $\tilde{a}=0$ are shown in Table \ref{finite temperature} and \ref{zero temperature}, where N represents having no ODLRO and O represents having ODLRO.  For $ \tilde{a}>0 $, at finite temperatures, ODLRO exists only in systems with $d=3$. So systems with dispersion $k_x^n+k_y^m+a (k_x^2+k_y^2)+k_z^2$ or $k_x^n+k_y^m+k_z^l+a (k_x^2+k_y^2+k_z^2)$ with $ l \geq m \geq n \geq 4 $ can be used to change the BEC into a noncondensed Bose liquid by tuning the shaking amplitude approaching the critical value $f_c$.

\section{COCLUSIONS} \label{conclusion}
In conclusion, we have investigated quantum phase transition of bosons in a shaken lattice by using Floquet theory and low-energy effective field theory. 
We found there was a $D_4$SF phase with spontaneous $D_4$ symmetry breaking and calculated the critical shaking amplitude $f_c$ for the NSF-$D_4$SF phase transition. We further demonstrated both the interaction effect induced by inhomogeneous band mixing and the shaking types could modify $f_c$. We identified a quantum tricritical point of NSF, $D_4$SF and MI phases and studied quantum criticality nearby the tricritical point. And the critical exponent is expected to be measured by \textit{in} \textit{situ} density measurements \cite{criticality exp} in the future. Moreover, we found anisotropically algebraic order and proposed to turn the BEC into a noncondensed Bose liquid by tuning the shaking amplitude approaching the critical value $f_c$.

\section*{ACKNOWLEDGEMENTS}
We thank H. Zhai, C. Chin, C. V. Parker and Q. Zhou for helpful discussions.

\end{document}